\documentclass[amssymb,amsmath,twocolumn,aps,groupedaddress,superscriptaddress]{revtex4}
\usepackage{epsfig}
\usepackage{graphicx}

\newcommand{\forget}[1]{}
\def\lambdabar{\protect\@lambdabar}
\def\@lambdabar{%
\relax
\bgroup
\def\@tempa{\hbox{\raise.73\ht0
\hbox to0pt{\kern.25\wd0\vrule width.5\wd0
height.1pt depth.1pt\hss}\box0}}%
\mathchoice{\setbox0\hbox{$\displaystyle\lambda$}\@tempa}%
{\setbox0\hbox{$\textstyle\lambda$}\@tempa}%
{\setbox0\hbox{$\scriptstyle\lambda$}\@tempa}%
{\setbox0\hbox{$\scriptscriptstyle\lambda$}\@tempa}%
\egroup
}

\forget{
\def\subequations{\refstepcounter{equation}%
\edef\@savedequation{\the\c@equation}%
\edef\@savedtheequation{\the\@stequation}
\edef\oldtheequation{\theequation}%
\setcounter{equation}{0}%
\def\theequation{\oldtheequation\alph{equation}}}%

\def\endsubequations{%
\setcounter{equation}{\@savedequation}%
\edef \theequation{\the\@steq} \global }
}

\begin{document}
\title{{Enhancing Acceleration Radiation from Ground-State Atoms
via Cavity Quantum Electrodynamics} }

\author{Marlan O. Scully}
\affiliation{Institute for Quantum Studies and Department of Physics, Texas A\&M Univ.,
TX 77843, USA}  \affiliation{Max-Planck-Institut f\"{u}r Quantenoptik, D-85748
Garching, Germany}
\author{Vitaly V. Kocharovsky}
\affiliation{Institute for Quantum Studies and Department of Physics, Texas A\&M Univ.,
TX 77843, USA} \affiliation{Institute of Applied Physics RAS, 603950 Nizhny Novgorod,
Russia}
\author{Alexey Belyanin}
\affiliation{Institute for Quantum Studies and Department of Physics, Texas A\&M Univ.,
TX 77843, USA} \affiliation{Institute of Applied Physics RAS, 603950 Nizhny Novgorod,
Russia}
\author{Edward Fry}
\affiliation{Institute for Quantum Studies and Department of Physics, Texas A\&M Univ.,
TX 77843, USA}
\author{Federico Capasso}
 \affiliation{Division of Engineering and Applied Sciences, Harvard University,
Cambridge, Massachusetts, USA}

\begin{abstract}
When ground state atoms are accelerated through a high Q microwave cavity, radiation is
produced with an intensity which can exceed the intensity of Unruh acceleration
radiation in free space by many orders of magnitude. The cavity field at steady state
is described by a thermal density matrix under most conditions. However, under some
conditions gain is possible, and when the atoms are injected in a regular fashion, the
radiation can be produced in a squeezed state.
\end{abstract}
\maketitle

One of the most intriguing results of modern quantum field theory is the proof by
Davies, Fulling, Unruh and DeWitt \cite{Davies}, and others \cite{2-7} that ground
state atoms, accelerated through vacuum, are promoted to an excited state just as if
they were in contact with a blackbody thermal field. These studies [1,2] predict that a
(two-level) ground state atom, having transition frequency $\omega$, and experiencing a
constant acceleration $a$, will be excited to its upper level with a probability
governed by the Boltzmann factor ${\rm{exp}} (-2\pi \omega /\alpha)$, where
$\alpha=a/c$, $c$ is the speed of light in vacuum. Unfortunately, even for very large
acceleration ``frequency" $\alpha$ $\approx10^{8}$ Hz \cite{8}, and microwave frequency
$\omega$ $\approx10^{10}$ Hz \cite{9}, this factor is exponentially small, $\sim
10^{-200}$; and is not of experimental interest.

\begin{figure}[ht]
\centerline{\includegraphics[scale=0.36]{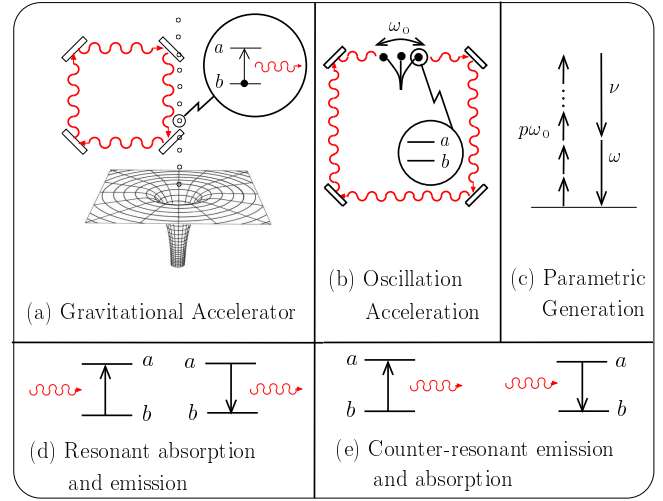}} \caption{\label{Fig01} (a) Atoms in
the ground state $|b\rangle$ are accelerated through small holes in the corner
reflectors of a microwave (or optical) cavity by, e.g., a strong gravitational field.
This is depicted as a unidirectional, single mode, ring cavity to convey the idea.
(b)``Vibrating reed" piezoelectrically driven oscillator containing a two-level atom is
placed in the cavity yielding strong mazer action . (c) Parametric conversion of
vibronic energy $p\hbar \omega_{{\rm{o}}}$ into photon and atom energies $\hbar \nu$
and $\hbar\omega$ respectively. (d) An atom is excited (de-excited) as it
simultaneously absorbs (emits) a photon in a resonant process. (e) The counter-resonant
processes that are usually neglected as compared to the resonant processes in the
``rotating wave" approximation; i.e. an atom is excited (de-excited) as it
simultaneously emits (absorbs) a photon.}
\end{figure}

Thus we were motivated to study a simple \textit{gedanken} experiment based on a model
consisting of a high Q ``single mode" cavity through which we pass accelerated
two-level atoms as in Fig. \ref{Fig01}. We find that the radiation is thermal (in the
typical case) and the effective ``Boltzmann factor" is now given by $\alpha/2\pi
\omega$. For the above example,  $\alpha /2\pi \omega \sim 10^{-3}$, hence, it is many
orders of magnitude larger than that for the usual Unruh effect and is potentially
observable.


The envisioned experiment can be described as a kind of ``acceleration radiation" mazer
\cite{Meyer,Agarwal}. In the ordinary maser, stimulated emission is the mechanism for
the production of radiation. In the present case, the physics of the emission process
is intimately association with the center-of-mass motion (taken in the $z$ direction).

One scheme for accelerating \cite{Yab} the atoms  uses a particle accelerator with,
e.g., hydrogen like ions. In such a case, ordinary (i.e. not Unruh) radiation emitted
by accelerated charged particles must be taken into account. Alternatively, we could
envision atoms accelerated in a strong gravitational field through a cavity. Other
means of operation via periodically driven atoms are also possible as in Figs. 1(b,c)
and are discussed later. For the moment, we simply assume the trajectories given by,
e.g., Eq. (2) and neglect the quantization of translational motion and recoil effect.

Our main results are contained in Eqs. (4)-(9). We find that the acceleration radiation
is generated by a kind of parametric process \cite{Boyd} in which both the atomic
polarization (the idler) and the radiation (the signal) are excited by extracting
energy from the atomic center-of-mass motion (the pump). Such processes are intimately
related to the so-called counter-rotating terms in the atom-field interaction
Hamiltonian and are discarded in the rotating wave approximation (RWA).

This provides a simple picture for the generation of acceleration radiation. The
photons emitted are real. The generation of radiation by the counter-rotating terms is
interesting; but, perhaps, no more bizarre than the earlier demonstration of mazer
emission \cite{Meyer} due to scattering of atoms off the cavity interface \cite{S.Z.}.
Furthermore, we find that the radiation may even be squeezed when $S_{1,2}$ in Eq. (8)
are nonvanishing. Calculation details will be given elsewhere \cite{ScullyTBP} as will
experimental implications \cite{TBP}.

As in the quantum theory of the laser \cite{Lamb,micromaser}, the (microscopic) change
in the density matrix of a cavity mode due to any one atom, $\delta \rho^{i}$, is
small. The (macroscopic) change due to $\Delta N$ atoms is then $\Delta \rho = \sum_i
\delta\rho^{i} = \Delta N \delta\rho$. Writing $\Delta N = r \Delta t$, where $r$ is
the atomic injection rate, we have a coarse grained equation of motion: $\Delta \rho
/\Delta t = r \delta \rho$. The change $\delta \rho^i$ due to an atom injected at time
$\tau_i$ in the atomic rest frame is
\begin{eqnarray}
\delta \rho^i =- \frac{1}{\hbar^2} \int_{\tau_i}^{\tau_i+ T}
\int_{\tau_i}^{\tau_i+\tau^{\prime}} {\rm{tr_{\rm atom}}} \times
\\ \nonumber \times \left[\hat{V}(\tau^{\prime}),\left[\hat{V}(\tau^{\prime
\prime}), \rho^{atom}(\tau_i)\otimes \rho(t(\tau_i)) \right]
\right]d\tau^{\prime}d\tau^{\prime \prime},
\end{eqnarray}
where $T$ is the proper time of flight through the cavity and tr$_{\rm atom}$ denotes
the trace over atom states. The time $\tau$ is the atomic proper time, i.e., the time
measured by an observer riding along with the atom. The cavity proper time $t(\tau)$
and the atomic trajectory of the atom as it passes through the cavity, $z(\tau)$, are
given by \cite{Rindler}
\begin{equation}
t(\tau)= t_0 + \frac{1}{\alpha} \rm{sinh}(\alpha \tau), \quad
z(\tau)=\frac{c}{\alpha}\left[ \rm{cosh}(\alpha \tau) - 1 \right],
\end{equation}
where $t_0 = t(\tau = 0)$ is the moment of time in the laboratory (cavity) frame when
the atom starts its acceleration. The distinction between atomic and cavity field
proper times is important. It is most convenient to calculate $\delta \rho^i$ in the
atomic frame. In the case of a running wave with a wave vector ${\bf k},\; k_z={\bf
k}\cdot{\bf v}/v$, the atom-field interaction Hamiltonian in the atomic frame is given
by
\begin{eqnarray}
\hat{V}(\tau)=\hbar g(\tau) \left[\hat{a}_k e^{-i \nu t(\tau) +
ik_zz(\tau)}+\rm{h.c.}\right] \left[\hat{\sigma}e^{-i\omega\tau}+\rm{h.c.}\right].
\end{eqnarray}
Here $g(\tau)=\mu E^{\prime}/\hbar$ is the atom-field coupling frequency which depends
on the atomic dipole moment $\mu$ and the electrical field $E^{\prime}$ in the frame of
the atom. For simplicity, consider the case of the co-propagating atom and field,
$k_z=|{\bf k}|=\nu/c$, so that $E^{\prime} = \sqrt{(c-v)/(c+v)}E$. Since
$v=c~{\rm{tanh}}(\alpha \tau)$ for a uniformly accelerated particle, we have
$E^{\prime}={\rm{e}}^{-\alpha\tau} E$ and $g(\tau)=g{\rm{e}}^{-\alpha \tau}$. The
operator $\hat{a}_k$ is the annihilation operator for the running wave, while
$\hat{\sigma}$ is the atomic lowering operator. Inserting Eq. (3) into Eq. (1) and
using Eq. (2), we obtain \cite{ScullyTBP} the results given in Eqs. (4)-(8) below.

   In the case of random injection times, the equation of motion
for the density matrix of the field is

\begin{eqnarray}
d\rho_{n,n}/dt ~=&& -R_2
\left[(n+1)\rho_{n,n}-n\rho_{n-1,n-1}\right]\\
\nonumber &&-R_1 \left[n\rho_{n,n}-(n+1)\rho_{n+1,n+1}\right],
\end{eqnarray}
where $R_{1,2}$ are defined in the following. If $R_1>R_2$, there is a steady state
solution which is thermal \cite{Lamb}
\begin{subequations}
\begin{equation}
\rho_{n,n}=e^{-\hbar\nu
n/k_B{\cal{T}}_c}\left(1-{\rm{e}}^{-\hbar\nu/k_B{\cal{T}}_c}\right),
\end{equation}
\begin{equation}
{\bar{n}}
=\sum_nn\rho_{nn}=\frac{1}{{\rm{e}}^{\hbar\nu/k_B{\cal{T}}_c}-1},
\; {\rm{e}}^{-\hbar\nu/k_B{\cal{T}}_c}=\frac{R_2}{R_1},
\end{equation}
\end{subequations}
where an effective temperature of the field in the cavity is ${\cal{T}}_c =\hbar
\nu/k_B{\rm {ln}}\left[R_1/R_2\right]$. Thus, spontaneous emission of randomly injected
ground state atoms in the cavity results in thermal statistics of the mode excitation.
Note, that the thermal statistics of the atomic excitation in the standard Unruh effect
in free space is due to spontaneous emission into a vacuum field reservoir with a
continuous spectrum of modes.

Absorption and emission coefficients $R_{1,2}=r|gI_{1,2}|^2$ are determined by the
amplitude $g {\rm e}^{-i\nu/\alpha} I_{1,2}= -\frac{i}{\hbar} \int_{\tau_i}^{\tau_i+T}
V_{1,2} d\tau$ of the matrix elements $V_1=\langle a,0|\hat{V}|b,1\rangle$ and
$V_2=\langle a,1|\hat{V}|b,0\rangle$ of the interaction Hamiltonian (3), respectively.
In particular case $\tau_i = 0$ we find
\begin{subequations}
\begin{equation}
I_1(\omega)=\int_{{\rm{0}}}^T
{\rm{exp}}\left[i\frac{\nu}{\alpha}{\rm{e}}^{-\alpha\tau}+
i\omega\tau-\alpha\tau\right]d\tau.
\end{equation}
It is convenient to write this as
\begin{equation}
I_{1}(\omega)=\left[\int^T_{\tau^*} d\tau - \int_{\tau^*}^{{\rm{0}}} d\tau \right]
{\rm{exp}}\left[i\frac{\nu}{\alpha}{\rm{e}}^{-\alpha\tau}+
i\omega\tau-\alpha\tau\right],
\end{equation}
\end{subequations}

\noindent where $\tau^* = -\infty -i \pi/2\alpha$. We carry out the first integral by
changing the variable of integration to $x=-i(\nu /\alpha){\rm{e}}^{-\alpha\tau}$ and
assume that ${\rm{e}}^{-\alpha T} \approx 0$. In such a case the first integral is
proportional to the ordinary gamma function defined as
$\Gamma(z)=\int_{{\rm{0}}}^\infty {\rm{e}}^{-x}x^{z-1} dx$. The second integral may be
adequately approximated by integration by parts in the limit that $\frac{\alpha}{\nu}
\ll 1$ and $\frac{\alpha}{\omega} \ll 1$. We find

\begin{subequations}
\begin{equation}
I_1(\omega)
=\frac{i}{\nu}\left(\frac{\alpha}{\nu}\right)^{-i\frac{\omega}{\alpha}}{\rm{e}}^{\frac{\pi\omega}{2\alpha}}
\Gamma\left(1-\frac{i\omega}{\alpha}\right)
\end{equation}
$$~~~~~~- \frac{i{\rm{e}}^{i\frac{\nu}{
\alpha}}}{\nu-\omega}\left[1+{\rm{O}}\left(\frac{\alpha\omega}{(\nu-\omega)^2}
\right)\right].$$

The corresponding integral for the emission of radiation $I_2(\omega)$ is equal to
$I_1(-\omega)$. We proceed to calculate $R_1\propto |I_1(\omega)|^2$ and $R_2\propto
|I_2(\omega)|^2$ by noting that $\Gamma\left(1+i\omega /\alpha\right)
\Gamma\left(1-i\omega / \alpha \right) = \left(\pi \omega / \alpha\right) /
{\rm{sinh}}(\pi \omega / \alpha)$.

We find that in the limit $\nu \gg \omega \gg \alpha$ the
emission/absorption ratio is $R_2/R_1 \simeq
\alpha/(2\pi\omega)$, which is an enhancement by many orders of
magnitude as compared to the exponentially small value $R_2/R_1 =
\exp(-2\pi\omega/\alpha)$.

For arbitrary values of parameters, the absorption and emission amplitudes can be
calculated as
\begin{equation}
I_{1,2}(\omega)=\frac{i}{\nu}\left(\frac{\alpha}{\nu}\right)^{\mp
i\frac{\omega}{\alpha}}{\rm{e}}^{\pm \frac{\pi \omega}{2\alpha}} \left[\Gamma(z,u {\rm
e}^{-\alpha T})-\Gamma(z, u)\right],
\end{equation}
\end{subequations}

\noindent where $z=1\mp i\frac{\omega}{\alpha}$, $u = - i\frac{\nu}{\alpha}$e$^{-\alpha
\tau_i}$, and $\Gamma(z,u) = \int_u^{\infty} {\rm e}^{-x}x^{z-1}dx$ is the incomplete
gamma function.

   The above analysis clearly shows that the mechanism of the
field and atom excitation in cavity quantum electrodynamics is the same as for the
Unruh effect in free space and is nothing but a nonadiabatic transition due to the
counter-rotating term $\hat{a}_k^+ \hat{\sigma}^+$ in the interaction Hamiltonian (3),
i.e. $V_2$. The reason for an enhanced excitation in the cavity is the relatively large
amplitude for a quantum transition $|b,0\rangle \rightarrow |a,1\rangle$ due to the
sudden nonadiabatic switching on of the interaction. As a result of this rapid turn on,
the initial state $|b,0\rangle$ is no longer an eigenstate of the Hamiltonian. Now, a
linear superposition of the excited states of atom and field makes up the dressed [14]
ground state of the interacting system $ \psi_0 = |b,0\rangle -
\frac{g(\tau)}{\nu^{\prime} + \omega} |a,1\rangle$ as well as the dressed excited state
$ \psi_1 = |a,1\rangle + \frac{g(\tau)}{\nu^{\prime} + \omega} |b,0\rangle$.

In particular, the amplitude of the bare excited state $|a,1\rangle$ in $\psi_0$ is of
the order of $C \sim \mu E^{\prime}/\hbar(\omega+\nu^{\prime})$. The latter corresponds
to the atomic excitation probability $\rho_{aa}^{atom} = |C|^2 \sim |\mu
E^{\prime}/\hbar(\omega+\nu)|^2 \sim |gI_2|^2$. This can be also found directly from
the density matrix equation for the atom, via the atomic counterpart to Eq. (1) with a
trace over the photon states instead of the tr$_{\rm atom}$. This probability has the
same origin and value as the well-known Bloch-Siegert shift of a two-level atomic
transition [14], $\Delta\omega/\omega = (\mu E^{\prime}/\hbar(\omega+\nu))^2$, due to
counter-rotating terms in the interaction Hamiltonian.

Clearly, the second term in Eqs. (6b) and (7a) represents the contributions from
boundaries to the nonadiabatic transition amplitudes. In the absence of the boundary
contributions, the emission integral $I_2(\omega) = I_1(-\omega)$ in Eqs. (6) and (7)
becomes exponentially small $\sim \exp(-\pi\omega/\alpha)$ for the small parameter
$\alpha/2\pi\omega \ll 1$ since there are no stationary phase points in the integration
interval. The absorption integral $I_1$ does have a point of stationary phase when the
atomic frequency $\omega$ is brought into resonance with the field due to the
time-dependent Doppler shift of the mode frequency \cite{doppler}
$\nu^{\prime}(\tau)=\nu\exp(-\alpha\tau)$. This fact explains why the related
exponential factor effectively disappears from the absorption integral (7a), $|{\rm
e}^{\frac{\pi \omega}{2 \alpha}} \Gamma(1 - i\omega/\alpha)| \simeq
(2\pi\omega/\alpha)^{1/2}$, when $\alpha \ll 2 \pi \omega$. As a result, if there are
no edge effects, we obtain the same excitation factor
$R_2/R_1=\exp(-2\pi\omega/\alpha)$ as in the Unruh effect (in free space). This means
that in order to observe the standard Unruh result one has to extend the mode profile
$g(z)$ near the boundaries, i.e., eliminate nonadiabatic boundary contributions
corresponding to the second term in Eq. (6b).

The nonadiabatic nature of the Unruh effect can be demonstrated most clearly by
following explicit derivation of the Unruh factor as a probability of the nonadiabatic
transition \cite{vkk}  $\psi_0 \rightarrow \psi_1$ from the dressed ground state.
Indeed, the Shroedinger equation $i\hbar d\psi/d\tau = H \psi$ in the two-level case
$\psi = c_0\psi_0 + c_1 \psi_1$ yields $dc_1/d\tau + (i E_1/\hbar + \langle
\dot{\psi_1}|\psi_1\rangle) c_1 = - c_{0}\langle \dot{\psi}_{0}|\psi_1\rangle$. The
difference between the eigenenergies is, to the first order, $E_1 - E_0 = \hbar (\omega
+ \nu^{\prime})$. For small nonadiabatic coupling $ -\langle \dot{\psi_0}|\psi_1\rangle
= \frac{d}{d\tau}\left(\frac{g(\tau)}{\omega + \nu^{\prime}}\right) \ll \omega +
\nu^{\prime}$, the perturbation solution is $|c_1|^2 = |\int_{\tau_i}^{\tau} \exp[i
\int_{\tau_i}^{\tau^{\prime}} (\nu^{\prime} + \omega) d\tau'']
\frac{d}{d\tau^{\prime}}\left(\frac{g(\tau^{\prime})}{\omega + \nu^{\prime}}\right)\,
d\tau'|^2$. If we now make the assumption of an adiabatic switching (on and off) of the
interaction $g(\tau)$ as in standard Unruh effect treatments, then after integration by
parts the latter integral is reduced to the integral $I_2(\omega) = I_1(-\omega)$ in
Eqs. (6) but in the infinite limits, i.e. without edge effects. This yields the
standard Unruh factor $|c_1|^2 \propto \exp(-2\pi\omega/\alpha)$. This derivation
clearly shows the dramatic effect of boundary contributions leading to a large
amplitude $\sim g(\tau)/(\omega + \nu')$ of the atomic excited state $|a\rangle$. Only
if we eliminate the edge effects by adiabatic switching of the interaction, do we
retrieve the exponentially small excitation factor.

   The surprising result is that in the cavity, the excitation
factor $\exp(-\hbar\nu/k_B{\cal{T}}_c)\equiv R_2/R_1=\alpha/2\pi\omega$ is determined
by the first power of the same nonadiabaticity parameter $\alpha/2\pi\omega$. The
reason for this effect is the existence of a true resonance, i.e., a stationary-phase
point, in the absorption coefficient (the first term in the integral $I_1$ in Eqs. (6b)
and (7a)). As mentioned earlier, this yields a resonance between the atomic transition
frequency and the Doppler-shifted frequency of the field seen by the atom, $\omega +
\frac{d}{d\tau}(\frac{\nu}{\alpha}{\rm {e}}^{-\alpha\tau}) \simeq 0$, and is
responsible for the aforementioned effect.

Another surprise of the cavity acceleration radiation is squeezing. If the atoms are
injected at regular intervals of times, $t_{0i}= \pi m_i/\nu +t_{\phi}$, where $m_i$ is
an integer, all atoms have the same phase with respect to the cavity mode, $\Phi
=\sum_{j=1}^{\Delta N}e^{-2i\pi m_j-2i\nu t_{\phi}}/\Delta N = e^{-2i\nu t_{\phi}}$,
and instead of Eq. (4) we find
\renewcommand{\theequation}{\arabic{equation}a}
  \setcounter{equation}{7}
\begin{eqnarray}
 && \dot{\rho}_{n,n} =
-R_1\left[n\rho_{nn}-(n+1)\rho_{n+1,n+1}\right] \\
\nonumber && -R_2\left[(n+1)\rho_{n,n}-n\rho_{n-1,n-1}\right]
\\
\nonumber && +[-S_1\sqrt{(n+1)(n+2)}\rho_{n+2,n}
-S_2\sqrt{(n-1)n}\rho_{n,n-2} \\
\nonumber && +(S_1+S_2)\sqrt{(n+1)n}\rho_{n+1,n-1}+{\rm{h.c.}}].
\end{eqnarray}
In this case the analysis is similar to the analysis of a polarization injected laser
\cite{S.Z.}, and the radiation density matrix is far from being thermal due to
squeezing factors
\renewcommand{\theequation}{\arabic{equation}b}
  \setcounter{equation}{7}
\begin{eqnarray}
S_{1,2}=rg^2 \Phi {\rm e}^{-2 i \nu/\alpha}\int_{\tau_i}^{\tau_i+T} d\tau^{\prime}
\int_{\tau_i}^{\tau^{\prime}} d\tau^{\prime \prime} \times \\
\nonumber e^{i \frac{\nu}{\alpha}e^{-\alpha\tau^{\prime}}\mp i \omega \tau^{\prime}
-\alpha\tau^{\prime}} e^{i \frac{\nu}{\alpha}e^{-\alpha \tau^{\prime \prime}}\pm i
\omega \tau^{\prime \prime} -\alpha\tau^{\prime \prime}}.
\end{eqnarray}
\renewcommand{\theequation}{\arabic{equation}}
\setcounter{equation}{8}

It is also possible to implement a more powerful resonant
emission by ground state atoms in a cavity, e.g. when the center
of mass of the atom is oscillating as
$z(\tau)=z_{{\rm{o}}}\cos(\omega_{\rm{o}}\tau)$, Fig. 1(b). This
can be viewed as another example of mazer action. In such a case,
the density matrix of a cavity mode is again found to obey Eq.
(8) but now
\begin{equation}
R_{1,2}\cong\frac{rg^2}{(\gamma+\alpha)^2}J_{{\rm{p}}}\left(-kz_{{\rm{o}}}\right)J_{{\rm{p}}}\left(kz_{{\rm{o}}}\right)
,
\end{equation}
where p is an integer, $J_p(x)$ the Bessel function, $\gamma$ the effective atomic
decay rate; and the squeezing terms $S_{1,2}$ are governed by cross terms which go as
$(rg^2/(\gamma+\alpha)^2)J_{{\rm{p}}}J_{0}$. Since this is a resonant parametric
process, the absorption ($p = 0$, $\omega = \nu$) and emission ($p \neq 0$, $\omega +
\nu = p \omega_0$) coefficients (9) are larger than for counter-rotating interactions,
Eqs. (6) and (7a), by a resonant factor $[\nu/(\gamma+\alpha)]^2$. In this case,
parametric generation is possible \cite{gain} (see \cite{TBP} for details).

{\underline{Concluding remarks}}. Our simple model clearly demonstrates that the ground
state atoms accelerated through a vacuum-state cavity radiate real photons. For
relatively small acceleration $a < 2\pi\omega c$, the excitation Boltzman factor
${\rm{exp}}(-\hbar\nu/k_B{\cal{T}}_c)\sim \alpha / 2 \pi \omega$ is much larger than
the standard Unruh factor $\exp(-2\pi\omega/\alpha)$. The physical origin of the
 field energy in the cavity and of the real internal energy in
the atom is, of course, the work done by an external force driving the center-of-mass
motion of the atom against the radiation reaction force. Both the present effect (in a
cavity) and standard Unruh effect (in free space) originate from the transition of the
ground state atom to the excited state with simultaneous emission of photon due to the
counter-rotating term $\hat{a}_k^+\hat{\sigma}^+$ in the time-dependent Hamiltonian
(3). The enhanced rate of emission into the cavity mode comes from the second term in
Eqs. (6b) and (7a)--the nonadiabatic transition at the boundaries of the cavity;  the
standard Unruh excitation comes from the first term in Eqs. (6b) and (7a) -- the
nonadiabatic transition in free space due to the time dependence of the Doppler-shifted
field frequency $\nu^{\prime}=\nu {\rm e}^{-\alpha\tau}$, as seen by the atom  in the
course of acceleration.

The authors gratefully acknowledge the support from DARPA-QuIST, ONR, and the Welch
Foundation. We would also like to thank R. Allen, H. Brandt, I. Cirac, J. Dowling, S.
Fulling, R. Indik, P. Meystre, W. Schleich, L. Susskind, and W. Unruh for helpful
discussions.


\begin{thebibliography}{xx}
\bibitem{9}
The frequency $\omega$ is lower bounded (for a possible experiment) by cryogenic
technology and the requirement that the effect should not be obscured by ``hot" walls.
\bibitem{doppler}
The Doppler-shifted frequency of the field, as it is seen by the atom is $\nu^{\prime}
=d(\nu t(\tau)-k_zz(\tau))/d\tau = \nu {\rm{e}}^{-\alpha \tau}$. For the case of a
counter-propagating wave one has to change the sign of $\alpha$ in the
${\rm{e}}^{-\alpha\tau}$ and related Eq. (6).
\end{thebibliography}
\end{document}